\documentclass[aps,pra,twocolumn,superscriptaddress,longbibliography,nofootinbib,floatfix]{revtex4-2}

\usepackage{amsmath,amssymb,amsfonts,bm}
\usepackage{graphicx}
\usepackage{dcolumn}
\usepackage[colorlinks=true,citecolor=blue,linkcolor=blue,urlcolor=blue,filecolor=blue]{hyperref}

\newcommand{\Tr}{\operatorname{Tr}}
\newcommand{\br}{\bm{r}}
\newcommand{\bs}{\bm{s}}
\newcommand{\bn}{\bm{n}}
\newcommand{\FQ}{\mathbb{F}_Q}
\newcommand{\FC}{\mathbb{F}_C}

\begin{document}

\title{Improving the estimation of spin-bath parameters via initial correlations and a two-qubit probe}

\author{Anila Khan}
\affiliation{Department of Physics, Quaid-i-Azam University, Islamabad 45320, Pakistan}

\author{Huzaifa Amjad}
\affiliation{Department of Physics, Quaid-i-Azam University, Islamabad 45320, Pakistan}

\author{Rizwan Abbas}
\affiliation{School of Science \& Engineering, Lahore University of Management Sciences, Opposite Sector U, D.H.A, Lahore 54792, Pakistan}

\author{Ali Raza Mirza}
\email{ali.mirza@qau.edu.pk}
\affiliation{Department of Physics, Quaid-i-Azam University, Islamabad 45320, Pakistan}
\affiliation{National Centre for Physics, Islamabad 45320, Pakistan}

\begin{abstract}
We study how a second qubit can improve the estimation of spin-environment parameters when only one of the two qubits is measured. Both qubits are coupled to a common spin bath through the exactly solvable central-spin model, in which populations and coherences both evolve. Because the bath operator coupled to the qubits commutes with the bath Hamiltonian, the environment induces no effective interaction between the qubits. Without initial correlations or direct inter-qubit coupling, the two-qubit probe therefore performs exactly like a single qubit. When initial system-environment correlations are included, preparing the second qubit reshapes the environment statistics in a parameter-dependent way, and the measured qubit reads out this imprint. This improves the quantum Fisher information (QFI) for the coupling strength by factors of $3-4$ relative to a correlated single-qubit probe, and by more than an order of magnitude relative to an uncorrelated one. For temperature estimation, the improvement reaches a factor of about $8$. A direct Ising coupling does not help by itself, but the advantage persists in its presence. We also derive the classical Fisher information for arbitrary projective measurements and identify the measurement that saturates the QFI, which in general lies out of the equatorial plane of the Bloch sphere.
\end{abstract}

\maketitle

\section{Introduction}

Open quantum systems are central to the development of quantum technologies \cite{nielsen2002quantum,haroche2014exploring}. Every quantum system inevitably interacts with its environment, which leads to decoherence \cite{schlosshauer2007decoherence,breuer2002theory,zurek1982environment}, and a quantitative description of this process requires the knowledge of environment parameters such as the system-environment coupling strength and the environment temperature. A natural way to access these parameters is to couple a small, controllable quantum system (a quantum probe) to the environment and to perform measurements on the probe once its dynamics has encoded the properties of the environment \cite{brunelli2011qubit,brunelli2012qubit,neumann2013high,benedetti2014quantum,correa2015individual,elliott2016nondestructive,norris2016qubit,tamascelli2016characterization,streif2016measuring,benedetti2018quantum,cosco2017momentum,sone2017exact,salari2019quantum,razavian2019quantum,gebbia2020two,wu2020quantum,tamascelli2020quantum,gianani2020discrimination,degen2017quantum}. The ultimate precision of such an estimate is fixed by the quantum Fisher information (QFI) via the quantum Cram\'er-Rao bound \cite{helstrom1976quantum,braunstein1994statistical,fujiwara2001quantum,monras2006optimal,monras2007optimal,paris2009quantum,genoni2011optical,spagnolo2012phase,pinel2013quantum,chaudhry2014utilizing,chaudhry2015detecting,giovannetti2011advances}; maximizing the QFI therefore minimizes the error of the estimate.

Most studies of quantum probing assume that the probe and its environment are initially in a product state. This assumption is questionable when the probe interacts strongly with its environment, because the probe and environment then build up correlations before the probe state is prepared \cite{hakim1985quantum,haake1985strong,grabert1988quantum,smith1990application,karrlein1997exact,romero1997decoherence,lutz2003effect,banerjee2003general,van2004new,ban2009quantum,campisi2009fluctuation,uchiyama2010role,dijkstra2010non,smirne2010initial,dajka2010distance,zhang2010different,tan2011non,lee2012noncanonical,morozov2012decoherence,semin2012initial,chaudhry2013amplification,chaudhry2013role,chaudhry2014effect,reina2014extracting,zhang2015role,chen2016effects,de2017dynamics,halimeh2017weak,kitajima2017expansion,buser2017initial,majeed2019effect,mirza2021master,mirza2024role,austin2020geometric,shahid2026investigating}. Remarkably, such initial correlations can be a resource for parameter estimation: they can increase the QFI considerably \cite{ather2021improving,mirza2024impact,mirza2025effect}. In parallel, it has been shown that a two-qubit probe coupled to a common harmonic-oscillator environment, of which only one qubit is measured, can outperform a single-qubit probe by orders of magnitude \cite{mirza2024improving}; related two-qubit sensing protocols have also been proposed for ac fields \cite{abbas2025sensing}. In the harmonic-oscillator case the improvement originates from the environment-induced (indirect) interaction between the two qubits, which depends on the environment parameters and is imprinted on the measured qubit \cite{mirza2024improving}.

In many solid-state platforms, however, the dominant environment is not bosonic but consists of other spins, e.g., nuclear spins surrounding a quantum dot or the $^{13}$C and electronic spin baths of nitrogen-vacancy centers \cite{prokofev2000theory,cucchietti2005decoherence,hanson2008coherent,de2010universal,neumann2013high}. It is therefore natural to ask whether, and through which mechanism, a two-qubit probe can improve the estimation of spin-environment parameters. In this paper we answer this question with an exactly solvable model: one or two qubits coupled to a bath of spin-$1/2$ particles \cite{cucchietti2005decoherence,majeed2019effect,mirza2024role}. In this model the qubits possess both a level splitting and a tunneling term, so that unlike pure-dephasing models both the populations and the coherences of the probe evolve \cite{majeed2019effect}.

Our first finding is a structural one. Since the bath operator that couples to the qubits commutes with the bath Hamiltonian, the spin environment acts on the qubits as a static, though quantum-mechanically distributed, field; it does \emph{not} induce any indirect interaction between them. Consequently, if the initial state is a product state and the two qubits are not directly coupled, the reduced state of the measured qubit is exactly the same as that of a single-qubit probe, and the second qubit is useless. The mechanism behind the two-qubit advantage of Ref.~\cite{mirza2024improving} is thus absent here. Our second, and central, finding is that the initial system-environment correlations restore and in fact considerably enlarge the usefulness of the second qubit. When the probe is prepared by a projective measurement on the correlated equilibrium state, the preparation of the \emph{unmeasured} qubit reweights the statistics of the environment in a manner that depends on the coupling strength and on the temperature. The measured qubit is sensitive to these statistics, so the second qubit acts as a parameter-dependent ``state-preparation lever'' on the environment. We show that this leads to a QFI that is $3$-$4$ times larger than that of a correlated single-qubit probe when estimating the coupling strength, and up to about $8$ times larger when estimating the temperature in marked contrast with the harmonic-oscillator case, where the two-qubit scheme brings little benefit for thermometry \cite{mirza2024improving}. We also study the role of a direct Ising interaction between the qubits, and we derive the classical Fisher information (CFI) for an arbitrary projective measurement to identify the optimal measurement.

The paper is organized as follows. In Sec.~\ref{sec:model} we introduce the model and solve the dynamics of a single qubit and of two qubits, with and without initial correlations. In Sec.~\ref{sec:qfi} we derive the QFI. Sections~\ref{sec:g} and \ref{sec:T} are devoted to the estimation of the coupling strength and of the temperature, respectively, and Sec.~\ref{sec:optimal} to the optimal measurement. We conclude in Sec.~\ref{sec:conclusion}. Details of the exact solution, the derivatives entering the QFI, and the Fisher-information formulas are given in the Appendices.

\section{The model and its dynamics}\label{sec:model}

\subsection{Single two-level probe}

We first consider a single qubit (spin-$1/2$) interacting with a bath of $N$ spin-$1/2$ particles. The total Hamiltonian is $H=H_S+H_B+H_{SB}$, with ($\hbar=k_B=1$ throughout) \cite{cucchietti2005decoherence,majeed2019effect}
\begin{align}
H_S &= \frac{\varepsilon}{2}\sigma_z+\frac{\Delta}{2}\sigma_x, \label{eq:HS}\\
H_B &= \sum_{i=1}^{N}\frac{\varepsilon_i}{2}\sigma_z^{(i)}+\sum_{i=1}^{N}J_i\,\sigma_z^{(i)}\sigma_z^{(i+1)}, \label{eq:HB}\\
H_{SB} &= \frac{1}{2}\sigma_z\otimes\sum_{i=1}^{N}g_i\,\sigma_z^{(i)}\equiv\frac{1}{2}\sigma_z\otimes \mathcal{B}. \label{eq:HSB}
\end{align}
Here $\sigma_{x,z}$ are Pauli operators of the probe, $\varepsilon$ is its level splitting and $\Delta$ its tunneling amplitude; $\varepsilon_i$ is the level splitting of the $i$th environment spin, $J_i$ the nearest-neighbor interaction between environment spins, and $g_i$ the coupling between the probe and the $i$th environment spin. Since $[H_S,H_{SB}]\neq0$, the probe energy is not conserved and both populations and coherences evolve.

The operators $H_B$ and $\mathcal{B}$ are diagonal in the product basis $|n\rangle\equiv|n_1\rangle|n_2\rangle\cdots|n_N\rangle$ ($n_i=0,1$) of the environment spins, $\mathcal{B}|n\rangle=G_n|n\rangle$ and $H_B|n\rangle=E_n|n\rangle$, with
\begin{align}
G_n&=\sum_i(-1)^{n_i}g_i,\nonumber\\
E_n&=\sum_i\Big[(-1)^{n_i}\frac{\varepsilon_i}{2}+(-1)^{n_i+n_{i+1}}J_i\Big]. \label{eq:GnEn}
\end{align}
Inserting the completeness relation $\sum_n|n\rangle\langle n|=\mathbb{I}$, the total unitary time-evolution operator becomes (see Appendix~\ref{app:exact})
\begin{equation}
U(t)=\sum_{n}e^{-iE_nt}\,U_n(t)\otimes|n\rangle\langle n|,\qquad U_n(t)=e^{-iH_nt}, \label{eq:U}
\end{equation}
where $H_n=\frac{\zeta_n}{2}\sigma_z+\frac{\Delta}{2}\sigma_x$ is the effective Hamiltonian of the probe for the environment configuration $|n\rangle$, with $\zeta_n=\varepsilon+G_n$. Explicitly,
\begin{equation}
U_n(t)=\cos(\Omega_nt)-\frac{i\sin(\Omega_nt)}{\Omega_n}H_n,\quad \Omega_n=\tfrac{1}{2}\sqrt{\zeta_n^2+\Delta^2}. \label{eq:Un}
\end{equation}
For every environment configuration the probe therefore precesses with frequency $2\Omega_n$ about the axis $\hat{\bn}_n=(\Delta,0,\zeta_n)/(2\Omega_n)$.

We take the probe to be prepared in a pure state $|\psi\rangle$ with Bloch vector $\bs$. We consider two initial system-environment states. The usual \emph{uncorrelated} state is $\rho(0)=|\psi\rangle\langle\psi|\otimes e^{-\beta H_B}/Z_B$, with $\beta=1/T$ and $Z_B=\Tr_B e^{-\beta H_B}$. The \emph{correlated} state is obtained by letting the probe and environment reach their joint equilibrium state $e^{-\beta H}/\Tr e^{-\beta H}$ and then preparing $|\psi\rangle$ by a projective measurement on the probe \cite{chaudhry2013amplification,chaudhry2013role,majeed2019effect,mirza2021master,ather2021improving}:
\begin{equation}
\rho(0)=|\psi\rangle\langle\psi|\otimes\frac{\langle\psi|e^{-\beta H}|\psi\rangle}{Z}, \label{eq:rho0c}
\end{equation}
with $Z=\Tr_{S,B}[|\psi\rangle\langle\psi|e^{-\beta H}]$.
In both cases the environment is diagonal in $\{|n\rangle\}$, and the Bloch vector $\br(t)=\Tr[\bm{\sigma}\rho_S(t)]$ of the probe takes the form (Appendix~\ref{app:exact})
\begin{equation}
\br(t)=\sum_n p_n\,\mathcal{R}_n(t)\,\bs, \label{eq:rt}
\end{equation}
i.e., a classical average of rigid rotations $\mathcal{R}_n(t)$ [Eq.~\eqref{eq:Rmatrix}] weighted by the probabilities $p_n$ of the environment configurations,
\begin{equation}
p_n^{\rm u}=\frac{c_n}{Z_B},\qquad p_n^{\rm c}=\frac{c_nA_n}{Z},\qquad c_n=e^{-\beta E_n}, \label{eq:pn}
\end{equation}
for the uncorrelated (u) and correlated (c) initial state, respectively. The factor
\begin{equation}
A_n=\langle\psi|e^{-\beta H_n}|\psi\rangle=\cosh(\beta\Omega_n)-\frac{\zeta_n s_z+\Delta s_x}{2\Omega_n}\sinh(\beta\Omega_n) \label{eq:An}
\end{equation}
encodes the effect of the initial correlations, and $Z=\sum_nc_nA_n$ \cite{majeed2019effect}. The physical content of Eq.~\eqref{eq:pn} is transparent: the preparation of the probe changes the probability of each environment configuration from $c_n/Z_B$ to $c_nA_n/Z$. Importantly, $A_n$ depends on the coupling strength through $\zeta_n$, whereas $c_n$ does not.

For the probe initially polarized along $x$, $|\psi\rangle=|+\rangle$ ($\bs=\hat{\bm x}$), Eq.~\eqref{eq:rt} gives
\begin{equation}
\br(t)=\sum_np_n\begin{pmatrix}(\zeta_n^2\cos\theta_n+\Delta^2)/(4\Omega_n^2)\\ \zeta_n\sin\theta_n/(2\Omega_n)\\ \Delta\zeta_n(1-\cos\theta_n)/(4\Omega_n^2)\end{pmatrix}, \label{eq:rplus}
\end{equation}
with $\theta_n=2\Omega_nt$, and $A_n=\cosh(\beta\Omega_n)-\Delta\sinh(\beta\Omega_n)/(2\Omega_n)$.

\emph{Homogeneous environment.} For the numerical results we take identical environment spins, $g_i=g$, $\varepsilon_i=\varepsilon_B$, and $J_i=0$. All quantities then depend on $n$ only through the bath magnetization $m=N-2k$ ($k$ being the number of spins in state $|1\rangle$), so that $G_n\to gm$, $E_n\to\varepsilon_Bm/2$, and each sum over $2^N$ configurations reduces to a sum over $N+1$ values of $m$ with degeneracy $d_m=\binom{N}{k}$. In particular $Z_B=[2\cosh(\beta\varepsilon_B/2)]^N$. This reduction is exact.

\subsection{Two two-level probes}

We now consider two qubits, labeled $a=1,2$, interacting with the common spin environment \cite{majeed2019effect}:
\begin{align}
H=&\sum_{a=1}^2\left[\frac{\varepsilon_a}{2}\sigma_z^{(a)}+\frac{\Delta_a}{2}\sigma_x^{(a)}\right]+\chi\,\sigma_z^{(1)}\sigma_z^{(2)}\nonumber\\
&+H_B+\frac{1}{2}\left(\sigma_z^{(1)}+\sigma_z^{(2)}\right)\otimes\mathcal{B}, \label{eq:H2}
\end{align}
where $\chi$ is a direct Ising interaction between the qubits. Because $\mathcal{B}$ and $H_B$ are still diagonal in $\{|n\rangle\}$, the time-evolution operator retains the structure of Eq.~\eqref{eq:U},
\begin{equation}
U(t)=\sum_ne^{-iE_nt}\,U_n^{(12)}(t)\otimes|n\rangle\langle n|,\quad U^{(12)}_n=e^{-iH_n^{(12)}t}, \label{eq:U2}
\end{equation}
with $H_n^{(12)}=H_n^{(1)}+H_n^{(2)}+\chi\sigma_z^{(1)}\sigma_z^{(2)}$ and $H_n^{(a)}=\frac{\zeta_n^{(a)}}{2}\sigma_z^{(a)}+\frac{\Delta_a}{2}\sigma_x^{(a)}$, $\zeta_n^{(a)}=\varepsilon_a+G_n$. We prepare the product state $|\psi\rangle=|\psi_1\rangle\otimes|\psi_2\rangle$, either from the uncorrelated state or, as in Eq.~\eqref{eq:rho0c}, by a projective measurement on \emph{both} qubits after the two qubits and the environment have reached their joint equilibrium. In complete analogy with the single-qubit case, the two-qubit state is
\begin{equation}
\rho_{12}(t)=\sum_np_n^{(12)}\,U_n^{(12)}(t)|\psi\rangle\langle\psi|U_n^{(12)\dagger}(t), \label{eq:rho12}
\end{equation}
with $p_n^{(12),{\rm u}}=c_n/Z_B$ and $p_n^{(12),{\rm c}}=c_nA_n^{(12)}/Z_{12}$, where $A^{(12)}_n=\langle\psi|e^{-\beta H_n^{(12)}}|\psi\rangle$ and $Z_{12}=\sum_nc_nA_n^{(12)}$.

Following Ref.~\cite{mirza2024improving}, we perform measurements on qubit~1 only; the second qubit is an auxiliary part of the probe. The state of qubit~1 is $\rho_1(t)=\Tr_2\rho_{12}(t)$.

\emph{No direct coupling ($\chi=0$).} Then $U^{(12)}_n=U_n^{(1)}\otimes U_n^{(2)}$ and $A_n^{(12)}=A_n^{(1)}A_n^{(2)}$, where $A^{(a)}_n$ is given by Eq.~\eqref{eq:An} with $\zeta_n\to\zeta^{(a)}_n$, $\Delta\to\Delta_a$, and $\bs\to\bs^{(a)}$. The partial trace over qubit~2 then gives exactly
\begin{equation}
\br_1(t)=\sum_n\tilde{p}_n\,\mathcal{R}^{(1)}_n(t)\,\bs^{(1)}, \label{eq:r1chi0}
\end{equation}
with $\tilde{p}^{\rm u}_n=c_n/Z_B$ and $\tilde{p}^{\rm c}_n=c_nA_n^{(1)}A_n^{(2)}/Z_{12}$.
Equation~\eqref{eq:r1chi0} is the key result of this section, and two remarks are in order. (i) For the uncorrelated initial state, $\br_1(t)$ coincides \emph{exactly} with the Bloch vector of a single-qubit probe. The physical reason is that $[\mathcal{B},H_B]=0$: the environment acts on each qubit as a static field $G_n$ drawn from a fixed distribution, and generates no indirect interaction between the qubits. This is in stark contrast with a harmonic-oscillator environment, for which the indirect interaction $\Delta(t)$ of Ref.~\cite{mirza2024improving} enters the dynamics of the measured qubit. (ii) For the correlated initial state, the second qubit enters $\br_1(t)$ through the factor $A^{(2)}_n$, which multiplies the probability of each environment configuration. Since $A_n^{(2)}$ depends on $g$ (through $\zeta_n^{(2)}$) and on $T$, the preparation of the unmeasured qubit imprints additional information about the environment parameters onto the statistics seen by the measured qubit. In particular, the logarithmic derivative (score) of the configuration probabilities, $\partial_x\ln\tilde p_n=\sum_a\partial_x\ln A^{(a)}_n+\partial_x\ln c_n-\langle\cdot\rangle$, acquires a second contribution, which is exactly doubled for identical qubits.

\emph{Direct coupling ($\chi\neq0$).} Now qubit~1 becomes entangled with qubit~2, and the reduced dynamics of qubit~1 depends on the state of qubit~2 even without initial correlations. For each environment configuration, $H^{(12)}_n$ is a $4\times4$ matrix that we diagonalize exactly; the sum over configurations is then performed exactly (Appendix~\ref{app:chi}). For the special case $\Delta_2=0$ a closed form exists: qubit~1 then sees the field $\zeta_n^{(1)}+2\chi q$ with probability $|\langle q|\psi_2\rangle|^2$, $q=\pm1$ [Eq.~\eqref{eq:Delta20}].

We have verified all the above expressions against a brute-force diagonalization of the full Hamiltonian (qubits plus $N=4$ environment spins), finding agreement to machine precision.

\section{Quantum Fisher information}\label{sec:qfi}

The QFI quantifies the precision with which a parameter $x$ can be estimated from the state $\rho_x$: for $M$ repetitions, any unbiased estimator satisfies the quantum Cram\'er-Rao bound $\delta x\geq1/\sqrt{M\FQ(x)}$ \cite{helstrom1976quantum,braunstein1994statistical,paris2009quantum}. In terms of the spectral decomposition $\rho_x=\sum_k\lambda_k|\lambda_k\rangle\langle\lambda_k|$,
\begin{equation}
\FQ(x)=\sum_k\frac{(\partial_x\lambda_k)^2}{\lambda_k}+2\sum_{k\neq l}\frac{(\lambda_k-\lambda_l)^2}{\lambda_k+\lambda_l}\left|\langle\lambda_l|\partial_x\lambda_k\rangle\right|^2. \label{eq:QFIgeneral}
\end{equation}
In both of our schemes the measured object is a single qubit, $\rho=\frac{1}{2}(\mathbb{I}+\br\cdot\bm{\sigma})$. Its eigenvalues are $\lambda_\pm=\frac{1}{2}(1\pm|\br|)$, and Eq.~\eqref{eq:QFIgeneral} reduces to (Appendix~\ref{app:fisher}) \cite{zhong2013fisher}
\begin{equation}
\FQ(x)=|\partial_x\br|^2+\frac{(\br\cdot\partial_x\br)^2}{1-|\br|^2}, \label{eq:QFI}
\end{equation}
where the second term is absent for pure states ($|\br|=1$). Equation~\eqref{eq:QFI} holds for the single-qubit probe [$\br$ from Eq.~\eqref{eq:rt}] and for the two-qubit scheme [$\br=\br_1$ from Eq.~\eqref{eq:r1chi0} or Appendix~\ref{app:chi}]. Unlike the pure-dephasing case of Ref.~\cite{mirza2024improving}, all three Bloch components evolve here, and the first term in Eq.~\eqref{eq:QFI} contains the population ($r_z$) contribution.

From Eqs.~\eqref{eq:rt} and \eqref{eq:r1chi0}, the derivative of the Bloch vector is
\begin{align}
\partial_x\br&=\sum_n\left[\partial_xp_n\,\mathcal{R}_n\bs+p_n\,\partial_x(\mathcal{R}_n\bs)\right], \label{eq:dr}\\
\partial_xp_n&=p_n\Big(\partial_x\ln w_n-\sum_{n'}p_{n'}\partial_x\ln w_{n'}\Big), \nonumber
\end{align}
where $w_n$ denotes the unnormalized weight ($c_n$, $c_nA_n$, or $c_nA^{(1)}_nA^{(2)}_n$). The first term describes how the parameter changes the \emph{statistics} of the environment, the second how it changes the \emph{dynamics} for a fixed environment configuration. Two parameters are considered.

\emph{Coupling strength $g$.} Since $\partial_g\zeta_n=G_n/g$ ($=m$ for the homogeneous bath) and $c_n$ is independent of $g$,
\begin{equation}
\partial_g\ln w_n=\frac{G_n}{g}\sum_{a}\frac{\partial_{\zeta}A^{(a)}_n}{A^{(a)}_n},\qquad
\partial_g(\mathcal{R}_n\bs)=\frac{G_n}{g}\,\partial_\zeta(\mathcal{R}_n\bs), \label{eq:dg}
\end{equation}
where the sum runs over the qubits whose preparation is included in the weight [no term for the uncorrelated states, $a=1$ for the correlated single qubit, $a=1,2$ for the correlated two-qubit scheme]. Explicit expressions for $\partial_\zeta A_n$ and $\partial_\zeta(\mathcal{R}_n\bs)$ are given in Appendix~\ref{app:derivatives}. For uncorrelated states the configuration statistics are independent of $g$, and only the dynamical term survives.

\emph{Temperature $T$.} The rotations $\mathcal{R}_n$ do not depend on $T$, so only the statistical term contributes, with $\partial_T=-\beta^2\partial_\beta$ and
\begin{equation}
\partial_\beta\ln w_n=-E_n+\sum_a\frac{\partial_\beta A^{(a)}_n}{A^{(a)}_n}, \label{eq:dT}
\end{equation}
with $\partial_\beta A^{(a)}_n=\Omega^{(a)}_n\sinh(\beta\Omega^{(a)}_n)-\frac{1}{2}(\zeta^{(a)}_ns^{(a)}_z+\Delta_as^{(a)}_x)\cosh(\beta\Omega^{(a)}_n)$.
Equations~\eqref{eq:QFI}-\eqref{eq:dT} give the QFI in closed form. In the absence of initial correlations, the two-qubit scheme with $\chi=0$ yields exactly the single-qubit QFI, $\FQ^{\rm 2q,u}=\FQ^{\rm 1q,u}$, for every parameter, time and temperature.

\emph{Optimization and time window.} The QFI depends on the interaction time $t$. As in Ref.~\cite{mirza2024improving}, we maximize it over $t$ to obtain $\FQ^{\max}$. For identical couplings the precession frequencies of different configurations are (almost) equally spaced, by $\approx2g$, which produces recurrences of the coherence at $t_{\rm rec}\approx\pi/g$. Such recurrences are an artifact of perfectly homogeneous couplings and disappear for disordered couplings \cite{cucchietti2005decoherence}. We therefore maximize over $0<t\leq\tau$ with $\tau=4$, which is shorter than $t_{\rm rec}$ for all coupling strengths considered ($g\leq0.5$), so that only the initial decoherence window is used.

\emph{Parameters.} Unless stated otherwise we use $N=50$ environment spins, $\varepsilon_1=\varepsilon_2=2$, $\Delta_1=\Delta_2=1$, $\varepsilon_B=0.1$, $J_i=0$, and both qubits prepared in $|+\rangle$. A small $\varepsilon_B$ corresponds to an environment whose own Hamiltonian is weak compared to the probe-environment interaction, which is precisely the regime where the initial correlations matter most \cite{majeed2019effect,mirza2024role}. We use the following notation: $\FQ^{\rm 1q,u}$ and $\FQ^{\rm 1q,c}$ for the single-qubit probe without and with initial correlations, and $\FQ^{\rm 2q,u}$ and $\FQ^{\rm 2q,c}$ for the two-qubit scheme.

\section{Estimation of the system-environment coupling strength}\label{sec:g}

\begin{figure}[t]
\includegraphics[width=\columnwidth]{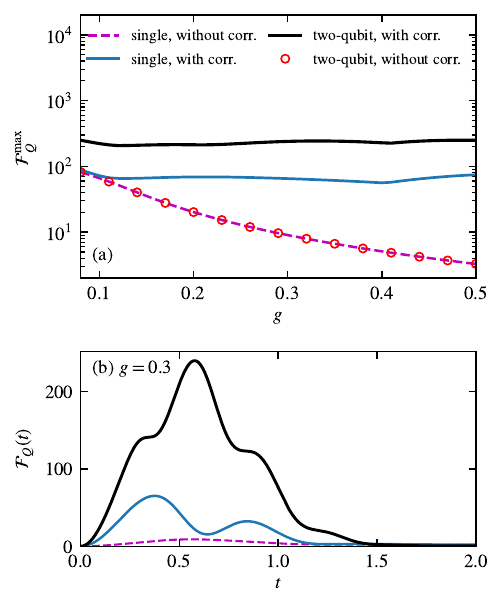}
\caption{Estimation of the system-environment coupling strength $g$ without direct inter-qubit interaction ($\chi=0$). (a) Optimized QFI versus $g$ for the single-qubit probe without (dashed magenta) and with (solid blue) initial correlations, and for the two-qubit scheme without (red circles) and with (solid black) initial correlations. The red circles lie exactly on the dashed magenta curve, as dictated by Eq.~\eqref{eq:r1chi0}. (b) QFI versus interaction time $t$ for $g=0.3$, with the same color code. Parameters: $N=50$, $\varepsilon_1=\varepsilon_2=2$, $\Delta_1=\Delta_2=1$, $\varepsilon_B=0.1$, $T=0.5$, both qubits prepared in $|+\rangle$, and the optimization in (a) is over $0<t\leq4$.}
\label{fig:g_chi0}
\end{figure}

We first estimate the coupling strength $g$ in the absence of a direct inter-qubit interaction, $\chi=0$. Figure~\ref{fig:g_chi0}(a) shows the optimized QFI as a function of $g$ at $T=0.5$. Several points should be noted. First, without initial correlations the two-qubit scheme (red circles) gives exactly the same QFI as the single-qubit probe (dashed magenta curve), as anticipated from Eq.~\eqref{eq:r1chi0}; in this scheme the second qubit is of no use. Second, the initial correlations are important already for a single qubit: $\FQ^{\rm 1q,c}$ (solid blue) stays between $56$ and $88$ over the whole range, while $\FQ^{\rm 1q,u}$ decreases monotonically, from $82$ at $g=0.08$ to $3.3$ at $g=0.5$, because stronger coupling means faster decoherence \cite{ather2021improving,mirza2024impact}. Third, and most importantly, with both the second qubit and the initial correlations taken into account (solid black), the QFI is increased by a factor between $2.8$ and $4.1$ relative to the correlated single qubit, and by a factor between $3$ and $76$ relative to the uncorrelated single qubit, the latter gain growing with the coupling strength. The improvement is not a mere doubling due to the use of two qubits: we find $\FQ^{\rm 2q,c}/(2\FQ^{\rm 1q,c})$ between $1.4$ and $2.0$, so measuring one qubit of the correlated two-qubit probe is more informative than measuring two independent, correlated single-qubit probes.

Figure~\ref{fig:g_chi0}(b) shows the time dependence of the QFI at $g=0.3$. The QFI vanishes at $t=0$, builds up as the environment imprints its properties on the probe, and decays once the probe has decohered; the correlated two-qubit scheme dominates during the whole decoherence window.

The origin of the improvement can be understood from the statistics of the environment. Consider the mean bath magnetization $\langle m\rangle=\sum_mp_mm$. For $g=0.3$ and $T=0.5$ we find $\langle m\rangle=-5.0$ for the uncorrelated state, independent of $g$; preparing a single correlated qubit shifts it to $\langle m\rangle=3.3$, and the additional preparation of the second qubit shifts it further to $\langle m\rangle=19.9$. Since the shift depends on $g$, the statistics of the effective field $G=gm$ experienced by the measured qubit carry information about $g$ beyond what is contained in its precession. The second qubit doubles this ``statistical'' part of the score [Eq.~\eqref{eq:dg}], which explains why the advantage is largest when the initial correlations are strong, i.e., at strong coupling and low temperature.

\emph{Robustness.} To check that our conclusions are not tied to the homogeneous couplings, we repeated the calculation for $N=12$ environment spins with disordered couplings $g_i=g(1+0.2\xi_i)$, $\xi_i$ drawn from a standard normal distribution, by summing exactly over all $2^{12}$ configurations. With the other parameters as in Fig.~\ref{fig:g_chi0}, we find $\FQ^{\rm 2q,c}/\FQ^{\rm 1q,c}=3.1$ and $3.7$ at $g=0.2$ and $g=0.5$, respectively, in close agreement with the corresponding homogeneous bath of $12$ spins ($3.0$ and $3.7$). The advantage also persists for other preparations of the auxiliary qubit; for example, for $g=0.3$ we obtain $\FQ^{\rm 2q,c}=240$, $235$, $303$, and $230$ for qubit~2 prepared in $|+\rangle$, $|-\rangle$, $|{\uparrow}\rangle$, and $|{\downarrow}\rangle$, respectively, compared with $\FQ^{\rm 1q,c}=65$.

\begin{figure}[t]
\includegraphics[width=\columnwidth]{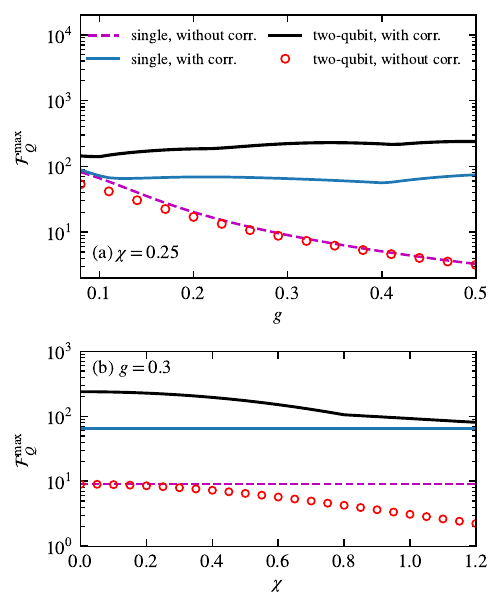}
\caption{Estimation of $g$ in the presence of a direct Ising interaction between the qubits. (a) Same as Fig.~\ref{fig:g_chi0}(a) but with $\chi=0.25$ for the two-qubit scheme (the single-qubit curves are unchanged). (b) Optimized QFI versus $\chi$ at $g=0.3$ for the two-qubit scheme without (red circles) and with (solid black) initial correlations; the horizontal lines show the single-qubit values without (dashed magenta) and with (solid blue) initial correlations. Other parameters as in Fig.~\ref{fig:g_chi0}.}
\label{fig:g_chi}
\end{figure}

\emph{Direct inter-qubit interaction.} We now switch on the Ising coupling $\chi$ between the qubits. Figure~\ref{fig:g_chi}(a) shows the results for $\chi=0.25$. Without initial correlations, the two-qubit scheme now differs from the single-qubit probe, but it performs \emph{worse}: $\FQ^{\rm 2q,u}/\FQ^{\rm 1q,u}$ lies between $0.65$ and $0.97$. The reason is that the coupling to qubit~2 shifts the field felt by qubit~1 by an amount that depends on the (undetermined) state of qubit~2. After qubit~2 is traced out, this acts as an additional source of dephasing, which carries little information about $g$ [for $\Delta_2=0$ it is a $g$-independent mixture of the shifts $\pm2\chi$; see Eq.~\eqref{eq:Delta20}]. A direct coupling is thus not a substitute for the environment-induced interaction of the harmonic-oscillator case \cite{mirza2024improving}. With the initial correlations included, however, the two-qubit scheme retains a clear advantage: $\FQ^{\rm 2q,c}$ exceeds $\FQ^{\rm 1q,c}$ by factors between $1.6$ and $3.9$, and $\FQ^{\rm 1q,u}$ by factors up to $73$. Figure~\ref{fig:g_chi}(b) shows how the optimized QFI depends on $\chi$ at $g=0.3$. Both two-qubit curves decrease with $\chi$, but the correlated two-qubit scheme remains above the correlated single-qubit probe over the whole range considered ($0\leq\chi\leq1.2$), decreasing from $240$ at $\chi=0$ to $81$ at $\chi=1.2$. Hence the best performance is obtained for $\chi=0$, and the advantage is robust against a moderate residual inter-qubit coupling.

\section{Estimation of temperature}\label{sec:T}

\begin{figure}[t]
\includegraphics[width=\columnwidth]{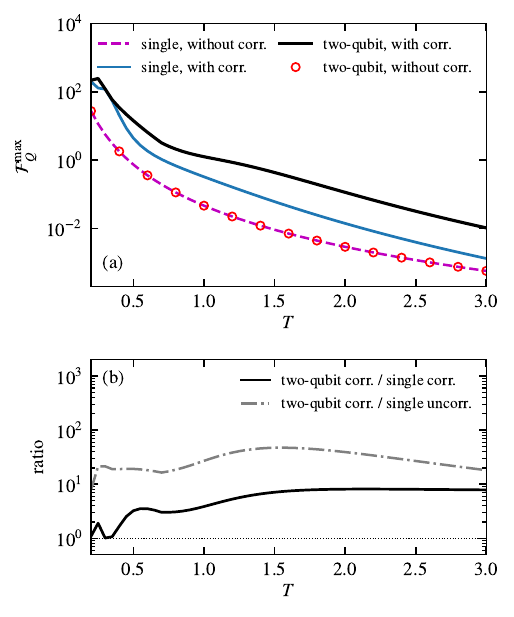}
\caption{Estimation of the environment temperature $T$ with $\chi=0$. (a) Optimized QFI versus $T$; color code as in Fig.~\ref{fig:g_chi0}(a). The red circles (two-qubit, uncorrelated) coincide with the dashed magenta curve (single qubit, uncorrelated). (b) Ratio of the optimized QFI of the correlated two-qubit scheme to that of the correlated (solid) and uncorrelated (dot-dashed) single-qubit probe. Here $g=0.3$; other parameters as in Fig.~\ref{fig:g_chi0}.}
\label{fig:T}
\end{figure}

We now turn to the estimation of the temperature $T$, with $\chi=0$ and $g=0.3$. The results are shown in Fig.~\ref{fig:T}. Without initial correlations, the probe is sensitive to $T$ only through the thermal configuration probabilities $c_n/Z_B$, and the two-qubit scheme again reproduces the single-qubit QFI exactly. With initial correlations, the factor $A_n$ adds a temperature dependence of its own and increases the QFI of the single-qubit probe by a factor of $2.3$-$21$ \cite{mirza2024impact,mirza2025effect}. The correlated two-qubit scheme further improves on the correlated single qubit by a factor that grows with temperature, from about $1$ at the lowest temperatures ($T\lesssim0.35$) to about $8$ for $T\gtrsim1.8$ [Fig.~\ref{fig:T}(b)]; relative to the uncorrelated single qubit, the gain lies between $8$ and $47$. The gain is weakest at very low temperature. There, the correlated preparation of both qubits drives the environment into a broad, nearly bimodal distribution of magnetizations (the standard deviation of $m$ is about $40$ at $T=0.25$, compared with $14$ for the correlated single qubit and $7$ for the thermal state), which causes rapid dephasing of the measured qubit and offsets the additional information carried by the reweighting.

This behavior contrasts with the harmonic-oscillator environment, where the two-qubit scheme brought little benefit for thermometry because the indirect inter-qubit interaction is temperature independent \cite{mirza2024improving}. Here the advantage arises from the state-preparation-induced reweighting of the environment, which is intrinsically temperature dependent [Eq.~\eqref{eq:dT}].

\section{Optimal measurement}\label{sec:optimal}

The QFI is the maximum information that can be extracted over all quantum measurements. We now identify a measurement that attains it. We consider a projective measurement on qubit~1 along the unit vector $\bn=(\sin\vartheta\cos\varphi,\sin\vartheta\sin\varphi,\cos\vartheta)$, with projectors $\Pi_\pm=\frac{1}{2}(\mathbb{I}\pm\bn\cdot\bm{\sigma})$. The outcome probabilities follow from the Born rule, $P(\pm|x)=\Tr(\Pi_\pm\rho)=\frac{1}{2}(1\pm\bn\cdot\br)$. The classical Fisher information (CFI) of this measurement is \cite{hall2000quantum,jacobs2014quantum}
\begin{align}
\FC(x)&=\sum_{k=\pm}\frac{[\partial_xP(k|x)]^2}{P(k|x)}=-\sum_{k=\pm}P(k|x)\,\partial_x^2\ln P(k|x)\nonumber\\
&=\frac{(\bn\cdot\partial_x\br)^2}{1-(\bn\cdot\br)^2}. \label{eq:CFI}
\end{align}
Equation~\eqref{eq:CFI} holds for both schemes and for any parameter. In terms of the angles, $\bn\cdot\br=\sin\vartheta(r_x\cos\varphi+r_y\sin\varphi)+r_z\cos\vartheta$, and similarly for $\bn\cdot\partial_x\br$.

We now maximize Eq.~\eqref{eq:CFI} over $\bn$. As shown in Appendix~\ref{app:fisher}, the maximum is reached for $\bn$ along the eigenbasis of the symmetric logarithmic derivative,
\begin{equation}
\bn_{\rm opt}=\frac{\bm{b}}{|\bm{b}|},\qquad \bm{b}=\partial_x\br+\frac{\br\cdot\partial_x\br}{1-|\br|^2}\,\br, \label{eq:nopt}
\end{equation}
i.e., $\vartheta_{\rm opt}=\arccos(b_z/|\bm b|)$ and $\varphi_{\rm opt}=\arg(b_x+ib_y)$. Substituting Eq.~\eqref{eq:nopt} into Eq.~\eqref{eq:CFI} gives
\begin{equation}
\FC(x)\big|_{\bn_{\rm opt}}=|\partial_x\br|^2+\frac{(\br\cdot\partial_x\br)^2}{1-|\br|^2}=\FQ(x), \label{eq:CFIopt}
\end{equation}
so this projective measurement is optimal. Since $\bm{b}$ depends on the parameter, the optimal measurement depends on its true value, as usual in local estimation theory; in practice it can be approached adaptively.

It is instructive to compare with the pure-dephasing case. If $\Delta=0$, $r_z$ is conserved; for a probe prepared in the equatorial plane, $r_z=0$, and Eq.~\eqref{eq:nopt} yields $b_z=0$, so the optimal measurement lies in the equatorial plane and only the azimuthal angle $\varphi$ needs to be tuned, as in Ref.~\cite{mirza2024improving}. With $\Delta\neq0$, as here, the environment also drives population dynamics, $r_z\neq0$, and the optimal direction is in general tilted out of the equatorial plane. For example, at $g=0.1$ and $T=0.5$ we find $\vartheta_{\rm opt}\approx128^\circ$ for the correlated single qubit and $\vartheta_{\rm opt}\approx108^\circ$ for the correlated two-qubit scheme; the best equatorial measurement then recovers only $66\%$ and $90\%$ of the respective QFI.

\begin{figure}[t]
\includegraphics[width=\columnwidth]{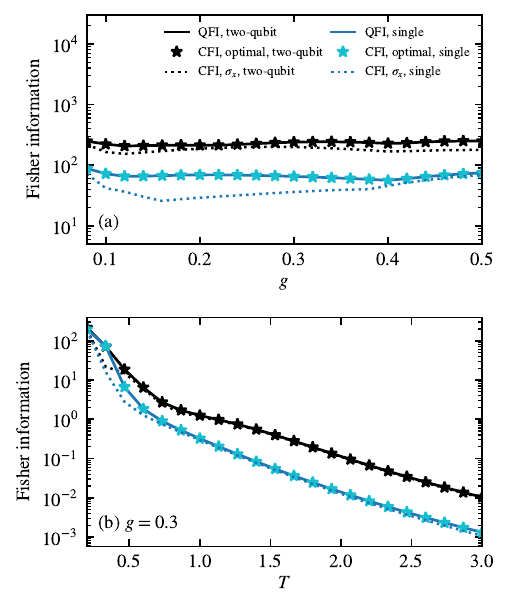}
\caption{Optimal measurement. Optimized QFI (solid lines), CFI of the optimal projective measurement of Eq.~\eqref{eq:nopt} (stars), and CFI of a fixed $\sigma_x$ measurement (dotted lines, separately optimized over $t$), for the correlated two-qubit scheme (black) and the correlated single-qubit probe (blue/cyan). (a) Estimation of $g$ at $T=0.5$. (b) Estimation of $T$ at $g=0.3$. Other parameters as in Fig.~\ref{fig:g_chi0}, with $\chi=0$.}
\label{fig:optimal}
\end{figure}

Figure~\ref{fig:optimal} confirms these results for the estimation of $g$ [panel (a)] and of $T$ [panel (b)], for the correlated single-qubit and two-qubit schemes. The CFI of the measurement in Eq.~\eqref{eq:nopt} (stars) coincides with the QFI (solid lines); the relative difference is below $10^{-15}$. For comparison, a fixed measurement of $\sigma_x$ (dotted lines), even when separately optimized over the interaction time, falls short of the QFI, most visibly for the single-qubit probe at intermediate coupling (where it captures only about $40\%$ of the QFI near $g\approx0.15$) and for temperature estimation near $T\approx0.3$. The improvement brought by the two-qubit scheme is thus accessible with a simple projective measurement on a single qubit.

\emph{Measuring both qubits.} By the monotonicity of the QFI under partial trace, measuring both qubits jointly cannot give less information than measuring qubit~1 alone; the values reported above are therefore lower bounds for the two-qubit probe. For instance, at $g=0.3$ and $T=0.5$ the QFI of the full two-qubit state $\rho_{12}$ is $356$ ($14$) with (without) initial correlations, compared with $240$ ($9.0$) when only qubit~1 is measured. We focus on the single-qubit readout because it requires only local measurements and allows a direct comparison with the single-qubit probe.

\section{Conclusion}\label{sec:conclusion}

We have studied the estimation of the parameters of a spin environment using a probe that consists of two qubits, of which only one is measured, and compared it with a single-qubit probe. Solving the central-spin model exactly, for one and two qubits and for uncorrelated and correlated initial states, we obtained closed-form expressions for the quantum Fisher information and for the classical Fisher information of an arbitrary projective measurement.

Our main conclusions are the following. (i) In the spin environment, unlike in a harmonic-oscillator environment, the bath does not mediate any interaction between the qubits; hence, without initial correlations and without direct coupling, the second qubit provides no advantage at all. (ii) Once the system-environment correlations present in the joint equilibrium state are taken into account, the preparation of the auxiliary qubit reshapes the statistics of the environment in a parameter-dependent way. This makes the two-qubit scheme $3$-$4$ times better than a correlated single-qubit probe for estimating the coupling strength, and up to about $8$ times better for estimating the temperature, with gains of more than an order of magnitude with respect to an uncorrelated single qubit. The advantage exceeds what would be obtained from two independent probes and survives disorder in the couplings. (iii) A direct Ising coupling between the qubits degrades the performance of the uncorrelated two-qubit scheme, but the advantage of the correlated scheme persists. (iv) The QFI can be saturated by a projective measurement on one qubit, whose direction we determined analytically; because the probe undergoes population dynamics, this direction is generally out of the equatorial plane.

These results identify initial system-environment correlations as the key resource that makes multi-qubit probes useful in spin environments. Natural extensions include multiparameter estimation, probes with more than two qubits for which the reweighting of the environment statistics should become even stronger and the combination of the present scheme with dynamical decoupling \cite{viola1998dynamical,viola1999dynamical,tan2014dephasing,shahid2026investigating}.

\begin{acknowledgments}
AR Mirza thanks the Higher Education Commission for support through IPFP (Interim Placement for fresh Phds) fellowship.
\end{acknowledgments}

\appendix

\section{Exact solution of the spin-spin model}\label{app:exact}

\subsection{Time-evolution operator}

Since $H_B$ and $\mathcal{B}$ commute and are both diagonal in the basis $\{|n\rangle\}$, we can write
\begin{align}
H\left(\mathbb{I}_S\otimes|n\rangle\langle n|\right)&=\left(H_S+\tfrac{1}{2}G_n\sigma_z+E_n\right)\otimes|n\rangle\langle n|\nonumber\\
&=(H_n+E_n)\otimes|n\rangle\langle n|. \nonumber
\end{align}
Therefore $e^{-iHt}=\sum_ne^{-iHt}(\mathbb{I}_S\otimes|n\rangle\langle n|)=\sum_ne^{-iE_nt}e^{-iH_nt}\otimes|n\rangle\langle n|$, which is Eq.~\eqref{eq:U}. As $H_n=\Omega_n\,\hat{\bn}_n\cdot\bm{\sigma}$ with $|\hat{\bn}_n|=1$, $(H_n)^2=\Omega_n^2$ and the exponential gives Eq.~\eqref{eq:Un}. In the Heisenberg picture $U_n^\dagger\bm{\sigma}U_n$ is the rotation of $\bm{\sigma}$ by the angle $2\Omega_nt$ about $\hat{\bn}_n$, so that a state with Bloch vector $\bs$ evolves into $\mathcal{R}_n(t)\bs$ with
\begin{equation}
\mathcal{R}_n(t)=
\left(\begin{array}{@{}c@{\,}c@{\,}c@{}}
\dfrac{\zeta_n^2\cos\theta_n+\Delta^2}{4\Omega_n^2} & -\dfrac{\zeta_n\sin\theta_n}{2\Omega_n} & \dfrac{\Delta\zeta_n\sin^2(\Omega_nt)}{2\Omega_n^2}\\[2mm]
\dfrac{\zeta_n\sin\theta_n}{2\Omega_n} & \cos\theta_n & -\dfrac{\Delta\sin\theta_n}{2\Omega_n}\\[2mm]
\dfrac{\Delta\zeta_n\sin^2(\Omega_nt)}{2\Omega_n^2} & \dfrac{\Delta\sin\theta_n}{2\Omega_n} & \dfrac{\zeta_n^2+\Delta^2\cos\theta_n}{4\Omega_n^2}
\end{array}\right), \label{eq:Rmatrix}
\end{equation}
$\theta_n=2\Omega_nt$. Equivalently, $\mathcal{R}_n\bs=\bs\cos\theta_n+(\hat{\bn}_n\times\bs)\sin\theta_n+\hat{\bn}_n(\hat{\bn}_n\cdot\bs)(1-\cos\theta_n)$. Equation~\eqref{eq:Rmatrix} agrees with the results of Ref.~\cite{majeed2019effect}.

\subsection{Uncorrelated initial state}

With $\rho(0)=|\psi\rangle\langle\psi|\otimes e^{-\beta H_B}/Z_B$ and $e^{-\beta H_B}=\sum_nc_n|n\rangle\langle n|$,
\begin{align}
\rho_S(t)&=\Tr_B\left[U(t)\rho(0)U^\dagger(t)\right]\nonumber\\
&=\frac{1}{Z_B}\sum_nc_n\,U_n(t)|\psi\rangle\langle\psi|U_n^\dagger(t), \label{eq:rhoSu}
\end{align}
because the phases $e^{\mp iE_nt}$ cancel in each diagonal block. Taking $\Tr[\bm{\sigma}\rho_S(t)]$ gives Eq.~\eqref{eq:rt} with $p_n=c_n/Z_B$.

\subsection{Correlated initial state}

Applying the same decomposition with $t\to-i\beta$ gives $e^{-\beta H}=\sum_nc_ne^{-\beta H_n}\otimes|n\rangle\langle n|$, so that the environment state in Eq.~\eqref{eq:rho0c} is
\begin{equation}
\rho_B(0)=\frac{\langle\psi|e^{-\beta H}|\psi\rangle}{Z}=\frac{1}{Z}\sum_nc_nA_n|n\rangle\langle n|, \label{eq:rhoBc}
\end{equation}
with $A_n=\langle\psi|e^{-\beta H_n}|\psi\rangle$.
Using $e^{-\beta H_n}=\cosh(\beta\Omega_n)-\sinh(\beta\Omega_n)H_n/\Omega_n$ and $\langle\psi|H_n|\psi\rangle=\frac{1}{2}(\zeta_ns_z+\Delta s_x)$ gives Eq.~\eqref{eq:An}. Note that $A_n\geq e^{-\beta\Omega_n}>0$. The rest of the calculation is identical to Eq.~\eqref{eq:rhoSu} with $c_n/Z_B\to c_nA_n/Z$. The initial environment state is diagonal but not thermal: its populations depend on the coupling strength and on the prepared probe state.

\subsection{Two qubits}

For two qubits, Eq.~\eqref{eq:H2} gives $H(\mathbb{I}_{12}\otimes|n\rangle\langle n|)=(H^{(12)}_n+E_n)\otimes|n\rangle\langle n|$ and hence Eqs.~\eqref{eq:U2} and \eqref{eq:rho12}. Preparing $|\psi_1\rangle\otimes|\psi_2\rangle$ by a projective measurement on both qubits gives the environment state $\rho_B(0)=Z_{12}^{-1}\sum_nc_nA^{(12)}_n|n\rangle\langle n|$. For $\chi=0$, $e^{-\beta H_n^{(12)}}=e^{-\beta H^{(1)}_n}\otimes e^{-\beta H^{(2)}_n}$, so $A_n^{(12)}=A^{(1)}_nA^{(2)}_n$, and $\Tr_2[U^{(1)}_n\otimes U^{(2)}_n\,\rho_1\otimes\rho_2\,U^{(1)\dagger}_n\otimes U^{(2)\dagger}_n]=U^{(1)}_n\rho_1U^{(1)\dagger}_n$, which leads to Eq.~\eqref{eq:r1chi0}.

For the homogeneous environment used in the figures, all sums run over $m=-N,-N+2,\dots,N$ with weights $d_m=\binom{N}{(N-m)/2}$, $G\to gm$, $E\to\varepsilon_Bm/2$, and $\zeta^{(a)}\to\varepsilon_a+gm$.

\section{Direct inter-qubit coupling}\label{app:chi}

For $\chi\neq0$, $H_n^{(12)}$ is a real symmetric $4\times4$ matrix. Writing $H^{(12)}_n=\sum_{j=1}^4\lambda_{n,j}|v_{n,j}\rangle\langle v_{n,j}|$, we have exactly
\begin{align}
|\psi_n(t)\rangle&=\sum_je^{-i\lambda_{n,j}t}\langle v_{n,j}|\psi\rangle|v_{n,j}\rangle,\\
A^{(12)}_n&=\sum_je^{-\beta\lambda_{n,j}}|\langle v_{n,j}|\psi\rangle|^2,
\end{align}
and $\br_1(t)=\sum_np^{(12)}_n\langle\psi_n(t)|\bm{\sigma}^{(1)}\otimes\mathbb{I}|\psi_n(t)\rangle$. For an auxiliary qubit without tunneling, $\Delta_2=0$, $\sigma_z^{(2)}$ is conserved and $H^{(12)}_n=\sum_{q=\pm1}|q\rangle\langle q|\otimes[H^{(1)}_n(\zeta^{(1)}_n+2\chi q)+\zeta^{(2)}_nq/2]$. The Bloch vector of qubit~1 is then
\begin{equation}
\br_1(t)=\sum_n\sum_{q=\pm1}p_n\,|\langle q|\psi_2\rangle|^2\,\mathcal{R}_{n,q}^{(1)}(t)\,\bs^{(1)}, \label{eq:Delta20}
\end{equation}
where $\mathcal{R}^{(1)}_{n,q}$ is Eq.~\eqref{eq:Rmatrix} with $\zeta_n\to\zeta^{(1)}_n+2\chi q$, and $p_n\propto c_n$ (uncorrelated) or
\begin{equation}
p_n\propto c_n\sum_{q=\pm1}|\langle q|\psi_2\rangle|^2e^{-\beta\zeta_n^{(2)}q/2}A^{(1)}_{n,q}\quad\text{(correlated)},
\end{equation}
with $A^{(1)}_{n,q}$ given by Eq.~\eqref{eq:An} with $\zeta_n\to\zeta^{(1)}_n+2\chi q$. Without correlations, qubit~1 thus sees a parameter-independent mixture of field shifts $\pm2\chi$, which can only reduce the QFI by convexity. We have checked Eq.~\eqref{eq:Delta20} against the exact $4\times4$ diagonalization.

\section{Derivatives entering the QFI}\label{app:derivatives}

For a probe prepared in $|+\rangle$, write $W_n=2\Omega_n=\sqrt{\zeta_n^2+\Delta^2}$ and $\theta_n=W_nt$. Differentiating Eq.~\eqref{eq:rplus} term by term,
\begin{align}
\partial_\zeta(\mathcal{R}_n\hat{\bm x})_x&=\frac{2\zeta_n\Delta^2(\cos\theta_n-1)}{W_n^4}-\frac{\zeta_n^3t\sin\theta_n}{W_n^3},\nonumber\\
\partial_\zeta(\mathcal{R}_n\hat{\bm x})_y&=\frac{\Delta^2\sin\theta_n}{W_n^3}+\frac{\zeta_n^2t\cos\theta_n}{W_n^2},\\
\partial_\zeta(\mathcal{R}_n\hat{\bm x})_z&=\frac{\Delta(\Delta^2-\zeta_n^2)(1-\cos\theta_n)}{W_n^4}+\frac{\Delta\zeta_n^2t\sin\theta_n}{W_n^3}.\nonumber
\end{align}
The terms proportional to $t$ describe the accumulation of phase, while the others arise from the change of the precession axis. For $A_n=\cosh(\beta\Omega_n)-\Delta\sinh(\beta\Omega_n)/(2\Omega_n)$, using $\partial_\zeta\Omega_n=\zeta_n/(4\Omega_n)$,
\begin{align}
\partial_\zeta A_n&=\frac{\zeta_n}{4\Omega_n}\Big[\beta\sinh(\beta\Omega_n)-\frac{\Delta\beta\cosh(\beta\Omega_n)}{2\Omega_n}\nonumber\\
&\qquad\qquad+\frac{\Delta\sinh(\beta\Omega_n)}{2\Omega_n^2}\Big],\\
\partial_\beta A_n&=\Omega_n\sinh(\beta\Omega_n)-\frac{\Delta}{2}\cosh(\beta\Omega_n).
\end{align}
These expressions, together with Eqs.~\eqref{eq:dr}-\eqref{eq:dT}, give $\partial_x\br$ in closed form. We have checked them against high-order finite differences (relative agreement better than $10^{-9}$).

\section{Fisher information of a qubit}\label{app:fisher}

\emph{QFI.} For $\rho=\frac{1}{2}(\mathbb{I}+\br\cdot\bm\sigma)$, the symmetric logarithmic derivative (SLD) $L$ is defined by $\partial_x\rho=\frac{1}{2}\{\rho,L\}$ and $\FQ=\Tr(\rho L^2)$. With the ansatz $L=a\mathbb{I}+\bm{b}\cdot\bm\sigma$ and $(\br\cdot\bm\sigma)(\bm b\cdot\bm\sigma)+(\bm b\cdot\bm\sigma)(\br\cdot\bm\sigma)=2\br\cdot\bm b$, the defining equation becomes $(a+\br\cdot\bm b)\mathbb{I}+(a\br+\bm b)\cdot\bm\sigma=\partial_x\br\cdot\bm\sigma$, whose solution is
\begin{equation}
a=-\frac{\br\cdot\partial_x\br}{1-|\br|^2},\qquad \bm b=\partial_x\br+\frac{\br\cdot\partial_x\br}{1-|\br|^2}\,\br.
\end{equation}
Then $\FQ=\Tr(\rho L^2)=a^2+|\bm b|^2+2a\,\br\cdot\bm b$, which simplifies to Eq.~\eqref{eq:QFI}. Equation~\eqref{eq:QFI} coincides with Eq.~\eqref{eq:QFIgeneral} evaluated with $\lambda_\pm=\frac{1}{2}(1\pm|\br|)$.

\emph{CFI and optimal measurement.} With $P(\pm|x)=\frac{1}{2}(1\pm\bn\cdot\br)$ we have $\partial_xP(\pm|x)=\pm\frac{1}{2}\bn\cdot\partial_x\br$ and $P(+|x)P(-|x)=\frac{1}{4}[1-(\bn\cdot\br)^2]$, which gives Eq.~\eqref{eq:CFI}. The SLD eigenbasis corresponds to $\bn\parallel\bm b$. Let $\alpha=\br\cdot\partial_x\br$ and $D=1-|\br|^2$. Then $\bm b\cdot\partial_x\br=|\partial_x\br|^2+\alpha^2/D$, $\bm b\cdot\br=\alpha/D$, and $|\bm b|^2=|\partial_x\br|^2+2\alpha^2/D+\alpha^2|\br|^2/D^2$, so that
\begin{equation}
\FC\big|_{\bn=\bm b/|\bm b|}=\frac{(\bm b\cdot\partial_x\br)^2}{|\bm b|^2-(\bm b\cdot\br)^2}=|\partial_x\br|^2+\frac{\alpha^2}{D}=\FQ,
\end{equation}
which proves Eq.~\eqref{eq:CFIopt}. Since $\FC\leq\FQ$ for every measurement \cite{braunstein1994statistical}, this measurement is optimal.

\bibliography{refs}

\end{document}